\documentstyle[twoside,fleqn,espcrc2,psfig]{article}
\newcommand{\beq}{\begin{equation}}
\newcommand{\eeq}{\end{equation}}
\newcommand{\bed}{\begin{displaymath}}
\newcommand{\eed}{\end{displaymath}}
\newcommand{\bea}{\begin{eqnarray}}
\newcommand{\eea}{\end{eqnarray}}

\renewcommand{\b}{\beta}
\renewcommand{\a}{\alpha}

\newcommand{\m}{\mu}

\newcommand{\s}{\sigma}

\newcommand{\dg}{\dagger}
\newcommand{\non}{\nonumber}

\newcommand{\rf}[1]{(\ref{#1})}
\newcommand{\ra}{\rightarrow}

\newcommand{\AmS}{{\protect\the\textfont2
  A\kern-.1667em\lower.5ex\hbox{M}\kern-.125emS}}

\title{Asymptotic Scaling, Casimir Scaling, and Center Vortices}

\author{M. Faber\address{Inst. f{\"u}r Kernphysik, Tech. Univ.
Wien, A-1040 Vienna, Austria}, J. Greensite\address{The Niels Bohr Institute,
DK-2100 Copenhagen \O, Denmark}
\thanks{Talk presented by J. Greensite.  Work supported by 
Carlsbergfondet, and by the U.S. Department 
of Energy under Grant No. DE-FG03-92ER40711.},
\v{S}. Olejn{\'\i}k\address{Inst. of Phys., Slovak Acad. of Sci., 
SK-842 28 Bratislava, Slovakia} }

\begin{document}
 
\begin{abstract}
    We report on two recent developments in the center vortex theory of 
confinement:
 (i) the asymptotic scaling of the vortex density, as measured in Monte Carlo
 simulations; and (ii) an explanation of Casimir scaling and the adjoint string
 tension, in terms of the center vortex mechanism.
\end{abstract}

\maketitle

\section{Introduction}

   Center vortices in SU(2) lattice gauge theory are detected on thermalized
lattices by fixing to the maximal center gauge, in which the quantity
\beq
     R = \sum_x \sum_\m \Bigl(\mbox{Tr}[U_\m(x)]\Bigr)^2
\label{mcg}
\eeq
is maximized, and then mapping link variables onto a $Z_2$ configuration
via center projection
\beq
       U_\m(x) \ra Z_\m(x) = \mbox{signTr}[U_\m(x)]
\eeq
The excitations of a $Z_2$ lattice gauge field are $Z_2$ vortices, which we
term "projection-vortices" or "P-vortices."  A plaquette is pierced by a 
P-vortex
if, in the center-projected lattice, the plaquette has the value $-1$.  We then
define the "vortex-limited" Wilson loops $W_n(x)$ to be loops on the 
unprojected
lattice evaluated when exactly $n$ P-vortices, in the center-projected 
lattice, pierce the minimal
area of the loop. It is then found that: (i) P-vortices on the projected 
lattice
locate (thick) center vortices on the unprojected lattice, as deduced from the
asymptotic behavior of the ratios $W_n(C)/W_0(C) \ra (-1)^n$; 
(ii) no vortices correlates with 
no confinement, i.e.\ $\chi_0(I,J) \ra 0$; (iii) vortices, by themselves, 
account for
the full asymptotic string tension obtained by standard methods, with the
potential linear beginning at one lattice spacing.  These results, together
with further references, are collected
in ref.\ \cite{Obs}.  It was also found that, upon abelian projection, 
center vortices appear as monopole-antimonopole chains.  The action of monopole
cubes is almost entirely concentrated in plaquettes pierced by P-vortices, and
the (unprojected) action distribution of monopole cubes is not 
much different from any other cube pierced by a P-vortex, c.f.\ \cite{Zako}.
For another approach to vortices, cf.\ \cite{TK}.

    In this talk I would like to report on two further aspects of center
vortices: asymptotic scaling and Casimir scaling.

\section{Vortex Density: Asymptotic Scaling}

\begin{figure}
\centerline{\hbox{\psfig{figure=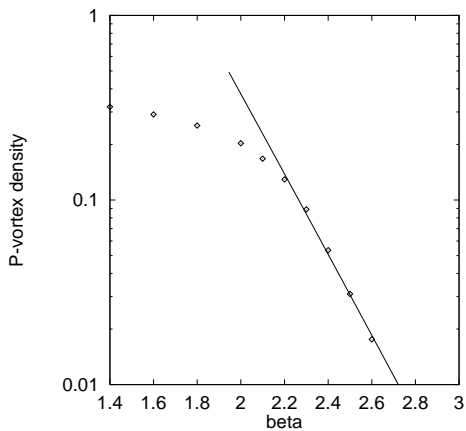,width=6.5cm}}}
\caption{Vortex density vs.\ coupling.}
\label{pvor}
\end{figure}

   The result that the density of center vortices scales according to the
asymptotic freedom prediction was first obtained by Langfeld et al.\ 
\cite{Langfeld}, using a slightly different version of maximal center gauge.
Here I will only display the results of our group, using the gauge obtained
by maximizing \rf{mcg}.  Define $N_{vor}$ to be the total number of 
plaquettes pierced by P-vortices, and $N_T$ the total number of plaquettes on
the lattice.  Then
\beq
      p = {N_{vor}\over N_T} = {\mbox{Total Vortex Area} \over 6
                  \times \mbox{Total Volume} } a^2 =
                   {1\over 6} \rho a^2
\eeq
where $\rho$ is the center vortex density (vortex area/unit volume) in physical
units.  Then, according to asymptotic freedom
\beq
     p = {\rho \over 6\Lambda^2} \left({6\pi^2 \over 11} \b \right)^{102/121}
            \exp\left[-{6\pi^2 \over 11}\b \right]
\eeq
Figure \ref{pvor} is a plot of $p$ vs.\ $\beta$. The straight line
is the asymptotic freedom result, with $\sqrt{\rho / 6\Lambda^2} = 50$.
Note that the scaling line has the slope appropriate to a
density of \emph{surfaces}.  The scaling lines for densities
of pointlike objects (instantons) or linelike objects 
(monopole loops) would have very different slopes.  
 
\section{Casimir Scaling}

    Our proposal is that the Casimir scaling of string tensions in the
intermediate distance regime is an effect due to the finite thickness of
center vortices \cite{Cas} (see also \cite{Corn}).  A model of those effects 
for SU(2) gauge theory, presented in more detail in
ref.\ \cite{Cas}, is the following:  If a vortex pierces the minimal area
of a planar Wilson loop, its effect is represented by the insertion of
a group element $G$    
\beq
     G = S \exp[i\a_C(x) \s_3/2] S^\dg
\eeq
into the product of (zero-vortex) link variables around loop C.  
It is assumed that
$S$ is a randomly distributed group element, and
$\a_C(x) \in [0,2\pi]$ depends on the location at which the vortex
pierces the minimal area of the loop.  If the cross-section of the vortex
``core'', where the vortex crosses the plane of the loop, is entirely
contained in the minimal area, then $G=-I$ ($\a=2\pi$), while if the
vortex cross section is entirely exterior to the minimal area, then
$G=+I$ ($\a=0$).  We have $G\ne \pm I$ if the vortex somewhere overlaps
the loop perimeter.  If we let $f$ denote the probability that the middle
of a vortex pierces any given plaquette, and further assume that the
positions of vortices piercing plaquettes in a plane are uncorrelated,
then the vortex contribution to the heavy-quark potential is found to
be
\bea
      V_j(R) &=& 
          - \sum_{n=-\infty}^{\infty}\ln\{(1-f)
          + f{\cal{G}}_j[\a_R(\mbox{x}_n)] \}
\non
 \\
     {\cal{G}}_j[\a] &=& 
         {1\over 2j+1} \sum_{m=-j}^{j} \cos(\a m)
\eea
where $j$ denotes the SU(2) group representation of the heavy quark
charges.  It can then be shown that the asymptotic string tension
is $\s_{j}= -\ln(1-2f)$ for $j=$half-integer, and $\s=0$ for 
$j=$integer.  For small $R$, on the other hand, where $\a_{R}(x)\ll 
2\pi$, we find
\beq
      V_j(R) = \left\{ {f\over 6}\sum_{n=-\infty}^{\infty} 
                 \a_R^2(\mbox{x}_n) \right\} j(j+1)
\eeq
which is proportional to the SU(2) Casimir (the result readily 
generalizes to SU(N)).  To go further, one needs some knowledge of
$\a_{R}(x)$.  The behavior of this function in the $R\ra 0$, $R\ra 
\infty$, and $x\ra \pm \infty$ limits is known, and most reasonable
ans{\"a}tze for $\a_{R}(x)$  do indeed display a Casimir-scaling region at
intermediate distances, where the potential is roughly linear, and
also roughly proportional to $j(j+1)$ \cite{Cas}.

\begin{figure}[h!]
\centerline{\hbox{\psfig{figure=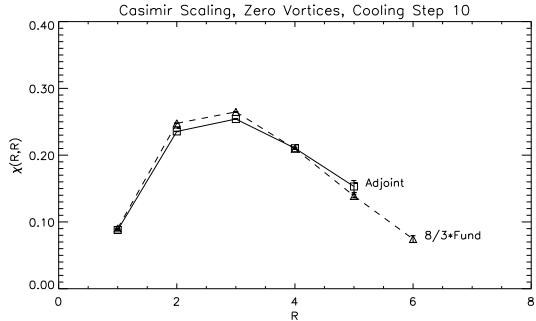,width=7.5cm}}}
\caption{Zero-vortex adjoint Creutz ratios, $\b=2.3$.}
\label{caszero}
\end{figure}

    If this explanation of Casimir-scaling is correct, then the string
tension of zero-vortex Wilson loops $W_{0}^{adj}(C)$ in the adjoint 
representation should disappear in the Casimir-scaling regime.  In
principle this is easy to test: we just compare Creutz ratios
$\chi_{0}^{adj}(I,J)$ with $\chi^{adj}(I,J)$.  In practice, however,
the VEV of adjoint loops is very small, and some reduction in noise
due to high-frequency fluctuations is essential.  This we have done
by applying the constrained cooling procedure of ref.\ \cite{cool}.
\begin{figure}[h!]
\centerline{\hbox{\psfig{figure=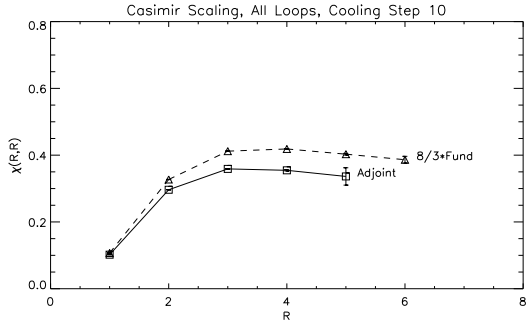,width=7.5cm}}}
\caption{Adjoint and fundamental($\times 8/3$) Creutz ratios,
$\b=2.3$.}
\label{casall}
\end{figure}
    Figure \ref{casall} is an illustration of Casimir scaling on the
cooled lattice ($\b=2.3$, 10 cooling steps).  The solid line shows Creutz 
ratios $\chi(R,R)$ in the adjoint representation (solid line), and in the 
fundamental representation rescaled by 8/3 (dashed line).  If Casimir 
scaling were exact, the two lines would coincide.  In both adjoint and
fundamental, the string tension seems to stabilize around $R=3$.  
Figure \ref{caszero} shows the corresponding quantities extracted from
zero-vortex Wilson loops.  In this case we notice that 
Casimir-scaling is nearly exact, and there is evidently no string 
tension; the Creutz ratios are clearly tending towards zero in both the
adjoint and fundamental representations.  It appears that when center
vortices are excluded from the interior of Wilson loops, both the 
fundamental \emph{and} adjoint string tensions disappear.  We view 
this as evidence in favor of a vortex origin of the adjoint string 
tension, in the Casimir-scaling regime.  Further details may be found
in ref.\ \cite{ECas}.

\section{Conclusions}

   The asymptotic scaling of the center vortex density strengthens 
our conviction that these surface-like objects identified by center
projection in maximal center gauge are physical objects.  Asymptotic
scaling is not 
the \emph{only} evidence of physicality; we have previously seen
that large Wilson loops linked to these objects pick up an extra factor of 
$-1$ \cite{Obs} (which allows us to identify them as center vortices), and 
that the vortices themselves are associated with
surfaces of anomalously high action-density \cite{Zako}. But 
certainly the scaling of the vortex density is an important part of 
the picture.

   We have also presented an explanation of the adjoint string tension,
in the context of the center vortex mechanism, and displayed some
numerical evidence that the existence of an adjoint tension correlates
with the presence of vortices linking the loop.  This provides, we 
believe, a plausible answer to the question ``What about Casimir Scaling?'' 
in connection with center vortices.

\end{document}